\documentstyle[preprint,aps]{revtex}

\begin{document}
\draft
\preprint{}
\title{The Magnetic Excitation Spectrum and Thermodynamics of
High-$T_c$ Superconductors}
\author{Pengcheng Dai,$^{1,\ast}$ H. A. Mook,$^{1}$ S. M. Hayden,$^{2}$ G.
Aeppli,$^{3}$
T. G. Perring,$^4$ R. D. Hunt,$^5$ and F. Do$\rm\breve{g}$an$^6$}
\address{\rm
$^1$Oak Ridge National Laboratory, Oak Ridge, Tennessee 37831-6393, USA.
$^2$H. H. Wills Physics Laboratory, University of Bristol, Bristol BS8 ITL, UK.
$^3$NEC Research Institute, Princeton, New Jersey 08540, USA.
$^4$Rutherford Appleton Laboratory, Chilton, Didcot, OX11 0QX, UK.
$^5$Oak Ridge National Laboratory, Oak Ridge, Tennessee 37831-6221, USA.
$^6$Department of Materials Science and Engineering,
University of Washington, Seattle, Washington 98195, USA. \\
$^\ast$ To whom correspondence should be addressed. E-mail: piq@ornl.gov
}
\maketitle
\begin{abstract}
Inelastic neutron scattering was used to study the wavevector- and
frequency-dependent magnetic fluctuations in single crystals of
superconducting YBa$_2$Cu$_3$O$_{6+x}$. The spectra contain several important
features, including a gap in the superconducting state, a pseudogap in the
normal state and the much-discussed resonance peak.
The appearance of the pseudogap determined from transport and nuclear
resonance
coincides with formation of the resonance in the magnetic excitations.
The exchange energy associated with the
resonance has the temperature and doping dependences as well as the magnitude to 
describe approximately the 
electronic specific heat near the superconducting transition temperature ($T_c$).
\end{abstract}
\newpage

\narrowtext
The parent compounds of the high-transition-temperature (high-$T_c$) cuprate
superconductors are antiferromagnetic (AF) insulators
characterised by a simple doubling of the crystallographic unit cells in the
CuO$_2$ planes. When holes are doped into these planes, the long-range AF
ordered phase
disappears and the lamellar copper oxide materials become metallic and
superconducting
with persistent short-range AF spin correlations (fluctuations).
Although spin fluctuations in cuprate superconductors
 are observed for materials such as YBa$_2$Cu$_3$O$_{6+x}$ [denote as
(123)O$_{6+x}$]
at all hole doping levels, $x$ \cite{mignod,mook,fong,dai1,fong1,bourges},
the role
of such fluctuations in the pairing mechanism and superconductivity
is still a subject of controversy \cite{conference}.
The most prominent feature in the fluctuations for (123)O$_{6+x}$ is a
sharp resonance
which for
highly  doped compositions appears below
the superconducting transition temperature $T_c$
at an energy of 41 meV \cite{mignod,mook,fong}.
We show that  the temperature-dependent resonance intensity is correlated
with the electronic part of the specific heat of (123)O$_{6+x}$,
$C^{el}(x,T)$ \cite{loram1,loram},
including the extended fluctuation regime (``pseudogap phase''), and that
the pseudogap
temperature $T^\ast$ determined
from the onset of the resonance intensity agrees with results of
transport and nuclear magnetic resonance (NMR) techniques.
By making absolute intensity measurements of the magnetic resonance,
$S_{res}(q,\omega)$, we estimate the contribution of its exchange energy
\cite{scalapino,demlerzhang} to the $C^{el}(x,T)$ anomaly around $T_c$.
We find that the temperature and doping
dependence of the resonance exchange energy can account for the 
$C^{el}(x,T)$ anomaly \cite{loram},
suggesting that a large part of this anomaly is due to
spin fluctuations.

Our experiments were performed using triple-axis spectrometers  at the
High-Flux
Isotope Reactor (HFIR) of the Oak Ridge National Laboratory
\cite{mook,dai1} and the MARI direct geometry
chopper spectrometer
at the ISIS pulsed spallation neutron source of the Rutherford Appleton
Laboratory \cite{aeppli}.
The momentum transfer $(q_x,q_y,q_z)$ is measured in units of \AA$^{-1}$ 
and reciprocal space positions are specified in reciprocal 
lattice units (rlu) $(h,k,l)=(q_xa/2\pi,q_yb/2\pi,q_zc/2\pi)$, 
where $a$, $b$ ($\simeq a$), and $c$ are the lattice parameters of the
orthorhombic unit cell of (123)O$_{6+x}$. For this study, we 
prepared four
single-crystal samples by the melt texture growth technique \cite{mook,dai1}.
Subsequent annealing resulted in oxygen stoichiometries of $x=0.6$, 0.7, 0.8,
0.93 with superconducting transitions at $T_c=62.7$, 74, 82, 92.5 K,
respectively.

The complete magnetic excitation spectra (Fig. 1) are obtained by
integration of the
magnetic neutron scattering over the two-dimensional
Brillouin zone (BZ) of (123)O$_{6+x}$ for $x=0.6$ at several temperatures
\cite{hayden,note1}. Because the CuO$_2$ planes in (123)O$_{6+x}$ actually
appear in coupled bilayers,
magnetic fluctuations
that are in-phase (acoustic or $\chi^{\prime\prime}_{ac}$) or out-of phase
(optical or $\chi^{\prime\prime}_{op}$) \cite{note2} with respect
to the  neighboring plane
will have different
spectra \cite{mignod,mook,fong,dai1,fong1,bourges}.
That the optical excitations occur at higher energies
than the acoustic excitations, and are characterized by a substantial 
gap even in
the normal state at 80 K,
demonstrates that the bilayer coupling remains antiferromagnetic on going
from the insulating parent to the
superconducting metal.
Upon lowering the temperature from 290 K to 35 K, the
most dramatic change in the local-susceptibility $\chi^{\prime\prime}(\omega)$
 for energies less than 100 meV is the appearance of a resonance
at 34 meV in the acoustic mode as shown in the shaded areas of Fig. 1.
For $x=0.6$, the resonance is clearly present not only below $T_c$ but also
at 80 K, albeit in
reduced form. Beyond the resonance, there is a broad feature extending at
least
to 220 meV, with a maximum at approximately 75 meV. This broad feature
is reminiscent of the continuum, peaked at around 20 meV,  seen in the
single layer compound
La$_{1.86}$Sr$_{0.14}$CuO$_4$ \cite{hayden1}. Well below the resonance
peak, scattering is
suppressed and a true spin-gap with a value of about 20 meV develops in the
superconducting state (Fig. 1A).
At room temperature, the optical and acoustic spectra (Figs. 1C and 1F)
show no maxima over the range of our measurements.

Because the most dramatic change in the magnetic spectrum is
the change in the resonance intensity, we focus on the
temperature and
composition dependence of the resonance.
The detailed momentum and frequency dependence of $S_{res}(q,\omega)$
obtained at HFIR at various temperatures around
the resonance energy for (123)O$_{6.6}$ (Fig. 2) show that
in the low temperature superconducting state (Fig. 2A), the
spectrum is dominated by the resonance at 34 meV.
At  temperatures just
above $T_c$ (Fig. 2B), the resonance broadens and decreases in intensity, consistent with
the ISIS data of Fig. 1. The resonance peak intensity appears to shift
smoothly toward
higher  energies in the normal state with increasing temperature.
A signature of the resonance remains at 125 K and 150 K, but essentially
vanishes at 200 K. Previous temperature dependent measurements
 for underdoped (123)O$_{6+x}$
indicate that the resonance peak
intensity changes its characteristics at $T_c$ \cite{dai1,fong1}. While the
intensity at
the resonance energy has a noticeable upturn upon cooling through $T_c$, the
scattering at frequencies above and below the resonance decrease below
$T_c$ \cite{dai1,fong1}.
This behavior can now be understood as arising from the narrowing of the resonance
in energy from the normal to the superconducting states.

The temperature dependence of the resonance peak intensity
[$S_{res}(q,\omega)$]
for three (123)O$_{6+x}$ samples with different doping levels and 
transition temperatures is shown in Fig. 3, A throught C.
For $x=0.6$,  the temperature dependence of the momentum- and frequency-
integrated resonance
is found to be different from that of the peak intensity (Fig. 3A). For
(123)O$_{6.8}$, similar behavior also
occurs,  but because the resonance is weaker above
$T_c$, the counting times required to obtain reliable integrals of the type
found for $x=0.6$ are prohibitive, and
so we are content with simply plotting the peak intensities. Finally,
for $x=0.93$, no broadening in either
energy or wave vector has been identified \cite{mignod,mook,fong}, and so
in this case, the temperature dependence
of the integrated spectral weight does actually follow the peak intensity.
We define the mean-squared
(fluctuating) moment associated with the resonance
as $
\langle m_{res}^2\rangle=3/(2\pi) \int d(\hbar\omega)\chi^{\prime\prime}_{res} 
(\omega)/(1-\exp(-\hbar\omega/kT))
$,
where $\hbar$ is the Planck's constant divided by $2\pi$, $k$ is the Boltzmann's
constant,
 $\chi^{\prime\prime}_{res}(\omega)$ \cite{note1,note2} is 
the resonance part of the acoustic spectrum of Fig. 1,
and the factor of
$1/2$ arises from averaging $\sin^2(q_z d /2)$ over $q_z$.
For (123)O$_{6.6}$ and (123)O$_{6.93}$ our measurements are expressed in 
absolute units obtained by
scaling to the low
temperature measurements performed on both compounds at ISIS \cite{hayden}.

The most obvious feature of Figs. 3, A throught C, is that as the doping level and
$T_c$ decrease,
there is a progressively larger pretransitional regime above $T_c$.
Specifically, for the ideally doped sample ($x=0.93$), the onset
of the resonance occurs at a temperature $T^\ast$ which almost coincides
with $T_c$. For $x=0.8$ and 0.6, $T^\ast$ increases to
approximately $115\pm 15$ and $150\pm 20$ K, respectively, even while both
$T_c$ and the resonance energy itself are
reduced. Thus, the weight of the temperature-dependent resonance 
joins the long
list of properties which show pretransitional
behavior in suboptimally doped (123)O$_{6+x}$. The cross-over temperature
$T^\ast$ (Fig. 4A) coincides with the temperatures
below which the temperature derivatives of the electrical resistivity
$d\rho(T)/dT$ \cite{ito,wuyts} and the Cu nuclear
$(T_1T)^{-1}$ relaxation rate \cite{takigawa} reaches a broad maxima. The
anomalies occurring at $T^\ast$ are generally
associated with the opening of a pseudogap in the low-energy spin excitation
spectrum, a supposition also supported by
neutron scattering data such as those in Fig. 1.

According to thermodynamics, a metal undergoes a transition into the superconducting 
state because such a transition can
lower its
total free energy, $F$ \cite{bcs}.
The difference in free energy of the system between the normal state,
extrapolated to zero temperature, ($F_N$) and
the superconducting state ($F_S$) is the condensation energy, that is,
$E_C=(F_N-F_S)_{T=0}$. In principle, the free energy of a system, and
therefore $E_C$, can be derived
from   the temperature dependence of $C^{el}(x,T)$ \cite{loram1}. For
(123)O$_{6+x}$, $C^{el}(x,T)$
has been measured by Loram {\it et al.} \cite{loram}.
The key features of the corresponding data (Fig. 3E)
are a sharp jump at $T_c$ for the optimally doped (123)O$_{6.92}$, which
becomes much suppressed upon reduction of the oxygen content.
Although there is less entropy released at $T_c$, more seems to be
released at temperatures above $T_c$, as $x$ is reduced below its optimal
value. Thus,
the specific
heat tracks the temperature derivative (Fig. 3D) of the spectral weight of
the resonance. Just as
for the specific heat,  decreasing doping reduces the maximum in $d\langle
m_{res}^2\rangle /dT$ at
$T_c$ and introduces progressively broader high-temperature tails.

In the $t$-$J$ model \cite{zhangrice}, 
the Hamiltonian of the system consists of a nearest-neighbor hopping 
($t$) term
accounting for the kinetic energy ($E_K$) of the carriers and a
magnetic exchange ($J$) term of the type describing the 
insulating antiferromagnetic parent compound \cite{hayden96b}.  
Magnetic fluctuations with a particular
wavevector
$q$ contribute to the exchange energy ($E_J$) via a product with $J(q)$, the
Fourier transform of the
exchange interactions \cite{scalapino}.
Although
magnetic spectra as complicated as those in
Fig. 1 will be difficult to  describe within such a simple model,
the exchange part of the $t$-$J$ Hamiltonian does
provide a convenient way to quantitatively estimate the effect of 
magnetic fluctuations on the thermodynamic properties of high-$T_c$
superconductors.
For (123)O$_{6+x}$, which has two coupled CuO$_2$ layers
per unit cell,
the exchange interactions consist of the nearest-neighbor spin-spin
coupling $J$
in the same CuO$_2$ plane and the coupling $J_\perp$ between two
CuO$_2$ planes within the unit cell. Since $J$ is much larger than $J_\perp$ \cite{hayden96b},
the exchange energy is \cite{scalapino,demlerzhang,noteeq}:
\begin{eqnarray}
E_J\simeq {3\over
(g\mu_{\rm B})^2}({a\over
2\pi})^2\int  \frac{d(\hbar\omega)}{\pi} \int dq_xdq_y 
\frac{1}{2}J[\cos(q_xa)+\cos(q_yb)]
\frac{[\chi^{\prime\prime}_{ac}(q_x,q_y,\omega)
+\chi^{\prime\prime}_{op}(q_x,q_y,\omega)]}
{1-\exp(-\hbar\omega/kT)}
\end{eqnarray} 
where $g$ is the Lande factor ($\simeq 2$) and $\mu_B$ is the Bohr magneton. Thus, 
Eq.~1 gives the contribution of
the exchange energy to the total energy.  Since the heat capacity is 
the temperature derivative of the total energy, we can estimate how 
the magnetic fluctuations responsible for the resonance affect the 
thermodynamic properties of (123)O$_{6+x}$ 
\cite{loram1,loram}. 
As the resonance grows, more spins fluctuating at the resonance frequency 
are correlated
antiferromagnetically, and the exchange energy is correspondingly reduced.
Assuming that the spins accounting for the resonance are not spatially correlated at high 
temperatures and therefore do not contribute to Eq. 1, the contribution
of the exchange energy of the resonance
($E_J^{res}$) to the specific heat, or $C^{res}_J(x,T)$, is
then
$$
C^{res}_J(x,T)\simeq {dE_J^{res}\over dT} =-{3\over 4\mu_{\rm B}^2}J  
\frac{d}{dT} \left( \int  \frac{d (\hbar\omega) }{\pi} 
\frac{\chi^{\prime\prime}_{res}(\omega)}{1-\exp(-\hbar\omega/kT)} 
\right)
 = -{J\over 2\mu_{B}^2} \frac{d \langle m_{res}^2\rangle}{dT} \;\;\;\; (2).
$$
Therefore,
$C^{res}_J(x,T)$ is approximately proportional to the temperature
derivative of
the spectral weight of the resonance, as we already discovered
from comparison of our
data with the measured specific heat. In deriving Eq. 2 from Eq. 1 we 
relied on the fact that
the resonance is sharply peaked around $(\pi/a,\pi/b)$.
The contribution of the resonance exchange energy to the heat capacity 
calculated using Eq. 2 with the measured exchange coupling $J=125\pm 20$ meV 
for (123)O$_{6.6}$ \cite{hayden} is shown in Fig. 3D (right
hand scale) and can be compared to the measured $C^{el}(x,T)$ of Loram {\it et al.} 
in Fig. 3E \cite{loram}. 
The model calculation with no
adjustable parameters shows that the resonance
can provide enough temperature-dependent exchange energy to yield the 
superconducting
anomaly in the specific heat \cite{note4}. This suggests that a large part
of the
specific heat near $T_c$ and its associated
entropy are due to spin fluctuations.

Clearly, the measured specific heat
includes contributions other than
that due to the exchange energy of the resonance ($E_J^{res}$).
While we included the most obvious
temperature dependent feature
in our calculation, the temperature evolution of the remainder of the spin
excitation spectrum is neglected. Most notably we ignored the formation 
of the spin pseudogap below $T^{\ast}$ and spin gap below $T_c$, 
both of which could 
increase $E_J$. 
Further, we have not considered the contribution of $E_K$
to $C^{el}(x,T)$. For superconductors, $E_K$ is expected to
increase from the normal to the
superconducting state \cite{scalapino}. Both this effect and spin (pseudo)gap  
formation could partially compensate for the 
reduction of $E_J$ due to the appearance of the resonance.
Such compensation might explain why the 
area of the calculated
$C^{res}_J(x,T)$ anomaly is considerably larger than that of the
$C^{el}(x,T)$ measurement. 

In the next few years, the contributions missing from the 
present analysis should become calculable using better neutron and optical 
conductivity data, which would provide the necessary information on the temperature-dependent 
magnetic exchange and kinetic energy terms, respectively. 
It would also be of considerable interest to make further model 
calculations to estimate the change in the resonance exchange energy between 
the normal and superconducting states at zero temperature and hence the
resonance exchange energy contribution to the condensation energy
\cite{scalapino,demlerzhang}. 
Unfortunately, this comparison is difficult because the normal state 
$S(q,\omega)$ must be extrapolated to zero temperature. This cannot be done
reliably with the present data and presents a challenge for the future.   

In summary, we show that a pseudogap regime bounded by a temperature
$T^\ast >T_c$
determined from transport, NMR, and thermodynamic measurements can be
associated with
the appearance of the resonance peak and the suppression of the low
frequency response in the
magnetic excitation spectrum.  Our complete spectra allow us to
establish the weight of the resonance relative to that of other spectral
features.
A simple calculation of the exchange energy using the measured
temperature dependence of the resonance
shows that
spin fluctuations can account for
a large part of the electronic 
specific heat near the superconducting transition.

\begin{figure}
\caption{Local- or wavevector-integrated frequency-dependent magnetic
susceptibility
$\chi^{\prime\prime}(\omega)$
for acoustic and optical modes in (123)O$_{6.6}$ corrected for
the isotropic Cu$^{2+}$ magnetic form factor and instrumental resolution [13]. 
The effect of Cu$^{2+}$ 
anisotropic form factor [S. Shamoto, M. Sato, J. M. Tranquada, B. J. Sternlieb,
and G. Shirane, Phys. Rev. B {\bf 48}, 13817 (1993)] is not considered.
 The data
were collected 
with the crystal mounted in the $(h,h,l)$ scattering plane
at MARI spectrometer at ISIS. ({\bf A}-{\bf C}) were chosen
at $q_z$ positions to emphasize acoustic modes and ({\bf D}-{\bf F}) the
optical modes.
Integrating the area under the resonance
peak and converting to moment squared yields $\langle m_{res}^2\rangle
= 0.06\pm0.04$ $\mu_{\rm B}^2$f.u.$^{-1}$ at 35 K and $0.04\pm0.03$
$\mu_{\rm B}^2$f.u.$^{-1}$
at 80 K. The resonance in (123)O$_{6.6}$ accounts for about 14\% of the
spectral weight up to 80 meV
at 35 K (Fig. 1A) and about 1\% of the total moment-squared
2$g^{2}\mu_{B}^{2}S(S+1)=6$ $\mu_{B}^2$ expected per formula unit (f.u.) 
assuming $S=1/2$ for each copper ion.
Similarly,  we find
that
 the 41 meV resonance for (123)O$_{6.93}$ has
$\langle m_{res}^2\rangle =0.06\pm0.04$  $\mu_{\rm B}^2$f.u$^{-1}$ at 25 K.
Note that the estimated error in $\langle m_{res}^2\rangle$
does not include the $\sim$30\%
systematic error involved
in normalizing the scattering to a vanadium standard.
}
\end{figure}

\begin{figure}
\caption{The magnetic structure factor $S_{res}(q,\omega$) around the
resonance energy for (123)O$_{6.6}$ at various temperatures. The data were
taken with the HB-3 triple-axis
spectrometer at HFIR using pyrolytic graphite as monochromator, analyzer,
and filters.
The
collimations were, proceeding from the reactor to the detector,
50$^\prime$-40$^\prime$-80$^\prime$-240$^\prime$, and the final neutron
energy was fixed at
30.5 meV. To compose the color figures,
constant-energy scans along the $(h,3h,1.7)$
direction [3,4] from $h=0.42$ to 0.58 rlu
were performed at energy transfers of $\hbar\omega = 25$, 27, 30, 34, 37,
40, 43,
and 46 meV. Magnetic intensity was
assumed to be scattering above the linear background of the
constant-energy scans. The scattering at different temperatures
was normalized to 300 monitor counts which correspond to 7 minutes per
point at $\hbar\omega = 34$ meV.
Note the change in the vertical color bar scales, made to
better display the dramatic decrease of the magnetic intensities on warming.
}

\end{figure}

\begin{figure}

\caption{
Temperature dependence of the
resonance peak intensity (filled circles) for ({\bf A}) (123)O$_{6.6}$ at
$\hbar\omega =34$ meV,
({\bf B}) (123)O$_{6.8}$ at 39 meV, and ({\bf C}) (123)O$_{6.93}$ at 40 meV
at the (0.5,1.5,1.7) rlu position on triple-axis spectrometers at HFIR.
Left axes show intensity normalized to $\langle m_{res}^2\rangle$ and right
axes to the 
peak intensity at low temperatures. The error bars in ({\bf A-C}) include only 
the statistical errors of the triple-axis measurements.  
The squares in ({\bf A}) are the integrated
$S_{res}(q,\omega)$ and the triangles in ({\bf C}) are from Ref. 2.
$T^\ast$s are defined as the initial appearance of the resonance.
The solid line in ({\bf C}) below $T_c$ is a fit using the modified two
fluid model [$I(T)/I(0)=1-(T/T_c)^{3.09}$].
({\bf D}) The estimated  $d\langle m_{res}^2\rangle /dT$ for (123)O$_{6+x}$
is from the temperature
derivative of the solid lines in ({\bf A}-{\bf C}). To calculate the
absolute magnitude of
$C^{res}_J(x,T)$  using Eq. 2, we have assumed $J=125$ meV (Ref. 13), $\langle
m_{res}^2\rangle =0.06$
$\mu_{\rm B}^2$f.u$^{-1}$, and 1 mole $=666.15$ grams for each of the three compositions.
Note we have not considered the uncertainties of $C^{res}_J(x,T)$ 
arising from the systematic errors in determining the absolute magnitude of  
$\langle m_{res}^2\rangle$, due most notably to the ambiguity in removing the 
nonresonance portion of the signal (Fig. 1A).  
({\bf E}) $C^{el}(x,T)$ from Ref. 9 was converted to SI units
using 1 g-at.$\equiv1/(12+x)$ moles.
}

\end{figure}

\begin{figure}
\caption{Phase diagram of (123)O$_{6+x}$ summarizing the results of
transport, NMR, and neutron scattering.
Because various methods of oxygenation, slightly different $T_c$ values are 
reported for the same nominal doping of (123)O$_{6+x}$.
We avoid this inconsistency by
plotting characteristic temperatures as a function of $T_c$. ({\bf A})
The open circles and open squares are temperatures which  $d\rho(T)/dT$
reaches a broad
maximum (Refs. 19, 20). The open diamonds show the the pseudogap
temperature $T^\ast$
determined from NMR measurements (Ref. 21).
The filled circles locate $T_c$ and $T^\ast$, where the resonance first appears
in our samples. ({\bf B})
Filled circles show resonance energy versus
the transition temperature $T_c$.
Filled squares are from Fong {\it et al.} (Ref. 5).
Horizontal error bars are superconducting transition
widths. The solid lines are guides to the eye.
}

\label{autonum}

\end{figure}


\begin{references}
\bibitem{mignod} J. Rossat-Mignod {\it et al.},
Physica (Amsterdam) {\bf 185C}, 86 (1991).
\bibitem{mook} H. A. Mook, M. Yethiraj, G. Aeppli, T. E. Mason, and T.
Armstrong,
Phys. Rev. Lett. {\bf 70}, 3490 (1993).
\bibitem{fong} H. F. Fong {\it et al.},
Phys. Rev. Lett. {\bf 75}, 316 (1995).
\bibitem{dai1} P. Dai, M. Yethiraj, H. A. Mook, T. B. Lindemer, and F.
Do$\rm\breve{g}$an, Phys. Rev. Lett. {\bf 77},
5425 (1996).
\bibitem{fong1} H. F. Fong, B. Keimer, D. L. Milius, and I. A. Aksay, Phys.
Rev. Lett. {\bf 78}, 713 (1997).
\bibitem{bourges} P. Bourges {\it et al.}, Phys. Rev. B {\bf 56}, R11439
(1997).
\bibitem{conference} Proceedings of the 10th Anniversary HTS Workshop on
Physics, Materials, and
Applications, Eds. B. Batlogg, C. W. Chu, W. K. Chu, D. V. Gubser, and K.
A. Muller, (World
Scientific) 1996.
\bibitem{loram1} J. W. Loram, K. A. Mirza, and P. F. Freeman, Physica C
{\bf 171}, 243 (1990).
\bibitem{loram} J. W. Loram, K. A. Mirza, J. R. Cooper, W. Y. Liang, and J.
M. Wade, J. Superconductivity {\bf 7}, 243
(1994).
\bibitem{scalapino} D. J. Scalapino and S. R. White, Phys. Rev. B {\bf 58},
8222 (1998).
\bibitem{demlerzhang} E. Demler and S. C. Zhang, Nature {\bf 396}, 733 (1998).
\bibitem{aeppli} See, for example, G. Aeppli, S. M. Hayden, and T. G.
Perring, Physics World,
December 1998, p33.
\bibitem{hayden} S. M. Hayden {\it et al.}, Physica B {\bf 241-243}, 765
(1998).
\bibitem{note1} We have used the elastic incoherent scattering
from a vanadium standard under the same experimental conditions to put the
two-dimensional wavevector integrated or local- susceptibility
$\chi^{\prime\prime}(\omega)=(a/2\pi)^2\int_{\rm
BZ}\chi^{\prime\prime}(q_x,q_y,\omega)dq_x dq_y$
from  (123)O$_{6+x}$
on an absolute scale (see, for example, Ref.\cite{hayden1}).
Although at low frequencies $\chi^{\prime\prime}(q,\omega)$
for (123)O$_{6.6}$ has been shown to peak at
wavevector positions incommensurate from the underlying crystallographic
lattice in the low temperature superconducting state \cite{dai2,mook1}, we
have only used a Gaussian convolved with the instrumental resolution 
to carry out
the
momentum integration within the scattering plane (see Fig. 6 of Ref.
\cite{hayden}). The poor out-of-plane
resolution of the instrument means that the integration of the magnetic
intensity over
momentum transfer perpendicular to the scattering plane is carried out
automatically.
The validity of such procedure to obtain $\chi^{\prime\prime}(\omega)$ has
been
confirmed by integrating the two-dimensional images of
$\chi^{\prime\prime}(q_x,q_y,\omega)$ at 24 meV and 34 meV \cite{mook1}.
\bibitem{note2}
It is convenient to split the magnetic response into two parts
$\chi^{\prime\prime}(q,\omega) =
\chi^{\prime\prime}_{ac}(q_x,q_y,\omega) \sin^2 (q_z d /2) +
\chi^{\prime\prime}_{op}(q_x,q_y,\omega) \cos^2 (q_z d /2)
$, where $d$ is the spacing between nearest-neighbor CuO planes.
\bibitem{hayden1} S. M. Hayden {\it et al.}, Phys. Rev. Lett. {\bf 76},
1344 (1996).
\bibitem{dai2} P. Dai, H. A. Mook, and F. Do$\rm\breve{g}$an, Phys. Rev.
Lett. {\bf 80}, 1738 (1998).
\bibitem{mook1} H. A. Mook {\it et al.}, Nature {\bf 395}, 580 (1998).
\bibitem{ito} T. Ito, K. Takenaka, and S. Uchida, Phys. Rev. Lett. {\bf
70}, 3995 (1993).
\bibitem{wuyts} B. Wuyts, V. V. Moshchalkov, and Y. Bruynseraede, Phy. Rev.
B {\bf 53}, 9418 (1996).
\bibitem{takigawa} M. Takigawa {\it et al.}, Phys. Rev. B {\bf 43}, 247 (1991).
\bibitem{bcs} J. R. Shrieffer, Theory of Superconductivity, {\it
Addison-Wesley, MA,} (1964).
\bibitem{zhangrice} F. C. Zhang and T. M. Rice, Phys. Rev. B {\bf 37}, 3759
(1988).
\bibitem{hayden96b} See, for example, S. M. Hayden, G. Aeppli, T. G. Perring, 
H. A. Mook, and F. Do$\rm\breve{g}$an, Phys. Rev. B {\bf 54},
R6905 (1996).
\bibitem{noteeq} Note that our susceptibility 
$\chi$ is defined as $\frac{1}{3}(\chi_{xx}+\chi_{yy}+\chi_{zz})$ while
$\chi_{+-}$ is used in  Ref.~\cite{demlerzhang}. We wish to thank
E. Demler for pointing this out and for many discussions about 
condensation energy in the high-$T_c$ superconductors.
\bibitem{note4} $C^{res}_J(x,T)$ is the resonance contribution
to the $C^{el}(x,T)$, not $C^{el}(x,T)$ itself. Therefore, one should only
compare $C^{res}_J(x,T)$ to the
change in $C^{el}(x,T)$ above the sloped background.
\bibitem{final} We thank B. Batlogg, E. Demler, R. S. Fishman, J. W. Loram,
D. Morr, D. Pines,
D. J. Scalapino, and S. C. Zhang for
stimulating discussions. The work at Oak Ridge
was supported by the US DOE under  Contract No. DE-AC05-96OR22464 with
Lockheed Martin
Energy Research Corp.
\end{references}
\end{document}